\documentclass{article} 

\usepackage{iclr_aims} 
\usepackage{titlesec}
\usepackage{natbib}

\iclrfinalcopy 

\usepackage[utf8]{inputenc} 
\usepackage[T1]{fontenc}    
\usepackage{hyperref}       
\usepackage{url}            
\usepackage{booktabs}       
\usepackage{amsfonts}       
\usepackage{nicefrac}       
\usepackage{microtype}      
\usepackage{times}          
\usepackage{graphicx}
\usepackage{amsmath,amssymb}
\usepackage{algorithmic}
\usepackage{textcomp}
\usepackage{xcolor}
\usepackage{tikz}
\usetikzlibrary{positioning,arrows.meta,shapes,fit}   
\titlespacing*{\section}{0pt}{0.5ex}{0.4ex}
\titlespacing*{\subsection}{0pt}{0.4ex}{0.3ex}
\titlespacing*{\subsubsection}{0pt}{0.3ex}{0.2ex}
\title{SOK: A Taxonomy of Attack Vectors and Defense Strategies for Agentic Supply Chain Runtime}

\author{Shiqi Yang$^{1,\dagger}$, Wenting Yang$^{1,\dagger}$, Xiaochong Jiang$^{1,\dagger}$, Yichen Liu$^{1}$, Cheng Ji$^{1}$ \\
$^1$Independent Researcher \\
\texttt{sy3506@nyu.edu}, \texttt{wey023@ucsd.edu}, 
\texttt{jiang.xiaoc@northeastern.edu}, \\
\texttt{yil160@ucsd.edu}, \texttt{chengji5@illinois.edu}
\thanks{$^\dagger$These authors contributed equally to this work.}
}

\begin{document}

\maketitle

\begin{abstract}
Agentic systems based on large language models (LLMs) operate not merely as text generators but as autonomous entities that dynamically retrieve information and invoke tools. This execution model shifts the attack surface from traditional build-time artifacts to inference-time dependencies, exposing agents to manipulation through untrusted data and probabilistic capability resolution. While prior work has examined model-level vulnerabilities, security risks arising from the complex, cyclic runtime behavior of agents remain fragmented.

This paper systematizes existing research into a unified runtime framework. We categorize threats into data supply chain attacks (distinguishing between transient context injection and persistent memory poisoning) and tool supply chain attacks (spanning discovery, implementation, and invocation phases). Crucially, we identify the emergence of the Viral Agent Loop, where agents effectively become vectors for self-propagating generative worms that require no code vulnerabilities to spread. We argue for a transition to a Zero-Trust Runtime Architecture, where context is treated as untrusted control flow, and tool execution is bounded by cryptographic provenance rather than semantic likelihood.
\end{abstract}

\vspace{-0.5em}
\vspace{1em}

\section{Introduction}
Large Language Models are increasingly deployed as autonomous agentic systems, moving beyond passive text generation. Modern agents retrieve information from external sources, maintain memory across interactions, and invoke tools capable of directly modifying digital or physical states \cite{wang2023voyager, xi2025rise}. As a result, agent behavior is influenced not only by model parameters and developer prompts, but also by dynamically acquired information and capabilities during execution.

This execution model establishes a fundamentally different security posture, referred to as Stochastic Dependency Resolution. Unlike traditional software systems, where dependencies such as \texttt{import numpy} resolve to a specific binary hash prior to deployment, agentic systems assemble their execution context at runtime based on semantic probability. Consequently, external documents, retrieved knowledge, APIs, and tools become implicit dependencies. Inference-time context therefore operates as an active component of the system’s attack surface, rather than serving as a passive input.

This shift challenges established cybersecurity assumptions. Autonomous agents routinely process untrusted data and execute privileged actions, often in the absence of direct human oversight. Consequently, attacks may occur without compromising infrastructure or model weights. Instead, adversaries can indirectly influence agent behavior by manipulating the environments with which agents interact, embedding malicious content into data sources or capabilities that agents may access during execution.

Existing research has explored several security risks associated with large language models, including training data poisoning, model backdoors, and framework-level vulnerabilities \cite{wang2024supplychain, zhao2025patronusidentifyingmitigatingtransferable, Liu_2026}. More recent work has identified specific attack vectors targeting agent behavior at runtime, such as indirect prompt injection during tool orchestration \cite{an2025ipiguard}, tool stream manipulation \cite{lin2026vigil}, and weaknesses in agent workflow auditing \cite{chen2025agentguard}. However, these efforts are typically studied in isolation, focusing on individual failure modes rather than the broader structure of agentic execution.

As agentic systems achieve greater autonomy, a fragmented perspective on security becomes insufficient. Runtime data ingestion, memory updates, and tool invocation are highly interdependent, and agent outputs may re-enter the system as subsequent inputs. These cyclic execution patterns enable persistent and self-reinforcing failures that isolated defenses cannot effectively mitigate. Addressing these risks requires a unified perspective that positions agent runtime behavior as a central cybersecurity concern.
\begin{figure*}[t]
\centering
\resizebox{0.7\linewidth}{!}{
\includegraphics[width=\linewidth,height=\textheight,keepaspectratio]{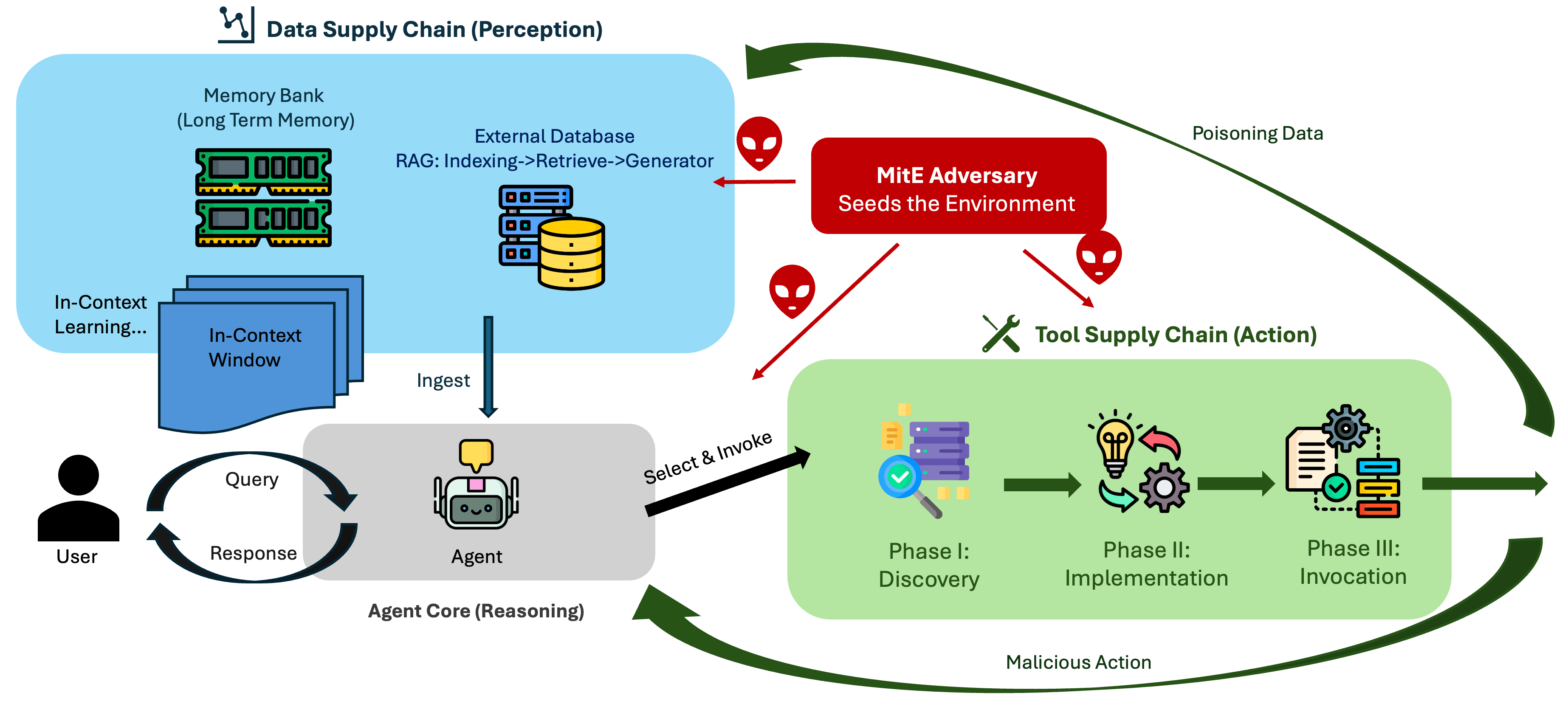}
}
\vspace{-2mm} 
\caption{The Agentic Runtime Supply Loop and associated attack surfaces.}
\label{fig:agent-loop}
\vspace{-2mm} 
\end{figure*}

This paper summarizes existing research on security threats to agentic systems arising from inference-time data and tool dependencies. By organizing prior work according to structural patterns in agent execution, the analysis clarifies relationships among diverse attack techniques and highlights shared security challenges in the existing agentic AI systems.

\section{Background and Threat Assumptions}
\label{sec:background}

Autonomous agentic systems differ from traditional software in how they acquire and trust dependencies.
Conventional applications rely on a \emph{static software supply chain}, where libraries and binaries are explicitly declared, resolved, and verified prior to deployment.
In contrast, agentic systems assemble their effective execution context at inference time: retrieved documents, external data sources, and tools—often selected autonomously—become implicit runtime dependencies that directly influence reasoning and action.

This shift fundamentally alters the security boundary.
Compromise may occur without modifying source code, model weights, or infrastructure, solely through manipulation of the external environments agents observe and interact with.
Consider the following canonical scenario of an Autonomous Travel Agent. Here, every execution step introduces an unvetted runtime artifact:

\noindent\fbox{%
    \parbox{0.95\columnwidth}{%
        \textbf{Running Example: Autonomous Travel Agent}\\
        An agent $\mathcal{G}$ is tasked with booking corporate travel.
        \begin{enumerate}
            \item \textbf{Goal:} Book a flight to a conference in Berlin and a nearby hotel.
            \item \textbf{Perception (Data SC):} $\mathcal{G}$ searches the web for ``Berlin conference hotels''.
            \item \textbf{Reasoning:} A retrieved blog claims a hotel offers a corporate discount via an external API.
            \item \textbf{Action (Tool SC):} $\mathcal{G}$ downloads and invokes a Python library suggested by the blog.
        \end{enumerate}
        Each step introduces runtime artifacts unknown and unvetted at deployment time, yet capable of influencing agent behavior.
    }%
}

\subsection{Static vs.\ Dynamic Supply Chains}

Traditional software systems rely on a finite, enumerable supply chain—source code, libraries, and binaries—resolved at build or deployment time.
Dependencies are explicitly declared, sourced from identifiable vendors, and amenable to pre-execution verification via signatures, audits, or vulnerability scanning.
Security mechanisms therefore focus on protecting infrastructure and validating static artifacts.


\begin{table}[htbp]
    \vspace{-0.5em} 
    
    \setlength{\abovecaptionskip}{2pt}
    \setlength{\belowcaptionskip}{0pt}

    \caption{\footnotesize Comparison: Static vs.\ Dynamic Supply Chains}
    \label{tab:supply_chain_comparison}
    \centering
    
    \resizebox{0.95\linewidth}{!}{%
        \begin{tabular}{@{}p{0.22\linewidth}p{0.34\linewidth}p{0.34\linewidth}@{}}
        \toprule
        \textbf{Feature} & \textbf{Static (Traditional)} & \textbf{Dynamic (Agentic)} \\ \midrule
        Resolution Time & Build / Deploy Time & Inference / Runtime \\
        Dependency Type & Libraries, Binaries & Data Context, APIs, Tools \\
        Topology & Directed Acyclic & Cyclic (Feedback Loops) \\
        Attack Surface & Code Vulnerabilities & Semantic Manipulation \\
        Upstream Source & Verified Vendor & Runtime Information Sources \\
        \bottomrule
        \end{tabular}%
    }
    
    \vspace{-0.5em} 
\end{table}

Agentic systems operate under a fundamentally different model.
Dependencies are acquired dynamically at runtime and selected through the agent’s reasoning process, producing a supply chain that is open-ended, non-enumerable, and environment-dependent.
As a result, attacks often operate through semantic manipulation of inference rather than binary exploitation of code.

\subsection{Threat Assumptions: Man-in-the-Environment (MitE)}We consider an adversary who does not compromise model weights, system prompts, or hosting infrastructure. Instead, the adversary exploits \textbf{Prompt--Data Isomorphism}, whereby retrieved passive data ($D$) is upcast into active instructions ($I$) during inference. We define the Man-in-the-Environment (MitE) adversary. In contrast to Man-in-the-Middle (MitM) attacks that compromise the channel, MitE exploits the agent's epistemic trust in the endpoint itself. By polluting the environment, the adversary corrupts the agent's ground truth. Because the agent treats the environment as its primary source of truth, the environment effectively becomes a malicious upstream vendor. In this model, the adversary acts as a \emph{runtime supplier}, injecting malicious artifacts into public data sources (wikis, repositories, forums) or tool registries. These artifacts are dormant until voluntarily retrieved and integrated by the agent, at which point they hijack the agent's reasoning loop.

\subsection{Adversary Capabilities and Goals}

We assume a \emph{runtime supplier} adversary capable of influencing agent behavior across both data and tool interfaces of the agentic supply chain.
The adversary may inject malicious content into data sources likely to be retrieved by the agent, steering reasoning via indirect prompt injection or related semantic manipulation.
Additionally, the adversary may publish deceptive tools or API wrappers that exploit the agent’s tool-selection logic, masquerading as legitimate functionality while executing unintended actions once invoked.

These capabilities enable several classes of harm.
An adversary may induce \emph{stealthy misalignment}, causing the agent to deviate from intended objectives without triggering safeguards, compromise confidentiality through information leakage during execution, or propagate malicious content back into the environment, enabling persistence or amplification as agent outputs are re-ingested as future inputs.

\section{The Data Supply Chain}
\label{sec:data_supply_chain}

This section explores agent's \textbf{ Perception Module}, the component of an agentic system responsible for acquiring, parsing, and normalizing external information from the environment including instruction prompts, historical dialogs, external knowledge database, etc. The attacker can inject malicious instructions or data into external information sources to subvert the agent’s subsequent reasoning and action. We categorize data-driven attacks based on where control is enforced, rather than on the surface form of the payload. Compared with standard LLM, the severity of attacks is amplified in agents through expanded \textbf{scope} and \textbf{persistence}. While traditional LLM data risks primarily result in ``one-off'' semantic corruption (e.g., misinformation or toxicity), agent-specific exploits target the \textbf{action-perception loop}. In an agentic workflow, the data supply chain is not merely a passive input stream; it functions as a functional \textbf{control signal} for tool-use and long-term planning. Consequently, a single poisoned data point does not yield simply a ``bad answer''---it induces a compromised operational state. This manifests itself through a shift from \textbf{static} to \textbf{dynamic} vulnerability: while standard LLM risks are often transient and limited to a single response, agentic risks can persist across sessions by prompt infection\cite{lee2024prompt}, poisoning the agent's self-updating memory\cite{zou2024poisonedrag} or triggering unauthorized high-stakes execution of external tools\cite{chen2024agentpoison}.

In this section, we categorize these data attacks based on the manipulation persistence: within-session manipulation and across-session manipulation.

\subsection{Within-Session Manipulation}
Within-Session Manipulation is transient and typically expire once the session is reset. This category targets the agent's \textbf{Contextual Window}, which is the immediate buffer of active tokens used for in-the-moment reasoning. 

\begin{enumerate}
\item \textbf{Indirect Prompt Injection:} The malicious payload is covertly embedded in a third-party, untrusted data source that the agent is instructed to process \cite{abdelnabi2023indirect}. The \textbf{mechanism} relies on the agent's \textbf{tool use} (e.g., a web browser or email API), which fetches the untrusted contents and prepends it to the user's query $Q$ before passing the entire context to the LLM. Furthermore, this vector extends beyond textual data into multi-modal domains: adversaries can embed subtle adversarial perturbations into \textit{images or sounds} that, upon processing, are decoded into a textual payload that steers the model to an attacker-chosen output (e.g., an image with an invisible patch that produces the instruction \texttt{``delete all files''}). This attack can lead to a 98\% success rate in steering model outputs\cite{bagdasaryan2023abusingimagessoundsindirect}.

\item \textbf{In-Context Learning}: These attacks exploit the mechanism of \textbf{In-Context Learning}(ICL)\cite{dong2023survey}. In the paper of \textbf{Many-Shot Jailbreaking} \cite{anil2024manyshot}, the attacker floods the short-term memory with a large volume of fictitious dialogues (e.g., 256+ turns) depicting the agent helpfully answering malicious queries. Due to the attention mechanism's prioritization of ICL over pre-trained safety guardrails, the agent ``learns'' from this poisoned short-term context that compliance with harmful requests is the correct behavior for the current session. Research indicates that as the shot count increases, the ASR follows a power-law growth, effectively rendering standard alignment filters obsolete within a single session.
\end{enumerate}

\subsection{Across-Session Manipulation}
This vector targets the integrity of the agent’s accumulated knowledge and historical context stored in \textbf{External Memory}. These attacks are persistent, allowing a single injection to influence the agent across multiple future sessions.

\begin{enumerate}
\item \textbf{Knowledge Base Contamination:} During Retrieval Aggregated Generation \textbf{retrieval}, the agent performs a semantic similarity search to extract relevant snippets $D'$ from the vectorized corpus, which are then integrated into the prompt to provide the LLM with grounded evidence. In the \textbf{generation} phase, the reasoning engine utilizes In-Context Learning to synthesize these external fragments into a coherent response or plan $P'$, a process that inadvertently treats the retrieved $D'$ as a high-priority instruction for the execution of the current task. For example, Zhong et al. \cite{zhong2023poisoning} showed that attackers can generate adversarial strings that hijack the vector space to ensure that they are retrieved for a wide range of indiscriminate user questions. Building on this, \textit{PoisonedRAG} injects strategically decomposed texts into a retrieval-optimized \textbf{trigger} and a generation-inducing \textbf{payload}. Studies show that poisoning only 0.1\% of the external corpus can result in 70\% ASR for targeted queries \cite{zou2024poisonedrag}.Approaches such as \textit{as AGENTPOISON} \cite{chen2024agentpoison} utilize constrained optimization to embed a stealthy backdoor. These backdoors remain dormant until a specific trigger in a future query activates the retrieval of the poisoned trace, with reported ASRs exceeding 80\% in various autonomous tasks \cite{chen2024agentpoison}. 

\item \textbf{Long Term Memory Poisoning:} Beyond static retrieval, another critical attack surface is the agent's \textbf{Long-Term Memory (LTM)}. Unlike the transient context window, LTM serves as a persistent repository that allows agents to store and retrieve episodic experiences (past trajectories) and semantic knowledge across different sessions \cite{weng2023memory}. This read-write autonomy enables the agent to accumulate ``experience,'' but it also creates a feedback loop vulnerability where the agent can be tricked into ``poisoning itself.'' Without having the privilege to the memory bank, \textit{MINJA} framework \cite{yao2025minja} proposed an attack mechanism that can inject malicious records into the memory bank only via queries. It appends an instruction-heavy 'indication prompt' to a benign query containing a specific 'victim term' to trigger the injection, achieving an average ASRs 76.8\%.
\end{enumerate}

\section{The Tool Supply Chain}
\label{sec:tool_supply_chain}

While the Data Supply Chain (Section~\ref{sec:data_supply_chain}) governs what information an agent incorporates into its internal state, the Tool Supply Chain governs what the agent can \emph{do} in the external environment. It binds natural-language intent to executable capabilities whose effects extend beyond the model boundary, directly modifying files, networks, accounts, or physical systems.

Failures in this chain therefore affect more than epistemic correctness. By corrupting the binding between \textbf{intent}, \textbf{capability}, and \textbf{authority}, an adversary can induce unintended or excessive actions while preserving apparently coherent reasoning traces. We refer to such failures as \textbf{capability hijacking}.

\paragraph{Tool Abstraction and Security Invariants.}
We model a tool as a delegated operational capability that binds semantic intent to executable code under bounded authority. A tool is characterized by its identifier, interface, authority scope, side effects, and output contract. Secure tool use relies on four core invariants: \textbf{identity integrity}, \textbf{semantic binding}, \textbf{authority bounding}, and \textbf{implementation integrity}. Violation of any single invariant is sufficient to induce externally observable harm.

\paragraph{Pipeline Decomposition.}
The Tool Supply Chain decomposes into three sequential capability-binding phases: Phase~I (\textbf{Discovery}) resolves intent to a concrete \texttt{Tool\_ID}; Phase~II (\textbf{Implementation}) fetches and instantiates executable code; and Phase~III (\textbf{Invocation}) executes the tool under a specific authority scope and side-effect envelope. Attacks may target any phase by corrupting tool identity, implementation integrity, or authority binding, with Phase~III operating closest to the execution boundary.


\subsection{Phase I: Attacks on Discovery (Intent $\rightarrow$ \texttt{Tool\_ID})}

The Discovery phase resolves natural-language intent into a concrete tool identifier prior to any code execution. In the benign setting, tool resolution follows:
\begin{equation}
f_{\text{resolve}}(\text{Intent}) \rightarrow \texttt{Tool\_ID}
\end{equation}

Under adversarial conditions, this resolution process may be perturbed by semantic noise or optimized manipulation of the surrounding context:
\begin{equation}
f_{\text{resolve}}(\text{Intent} + \epsilon_{\text{adv}}) \rightarrow \texttt{Tool\_ID}_{\text{malicious}}
\end{equation}
where $\epsilon_{\text{adv}}$ represents adversarially crafted semantic perturbations injected into tool descriptions, metadata, or retrieval context. 
Attacks at this stage compromise \emph{selection} rather than execution, manipulating how tools are indexed, described, or retrieved so that the agent voluntarily selects an adversary-controlled capability.

\begin{enumerate}
    \item \textbf{Hallucination Squatting:}
    During planning, agents may hallucinate plausible but non-existent tool identifiers. Adversaries can preemptively register these identifiers as malicious packages, transforming a would-be resolution error into successful execution of attacker-controlled code. Empirical evidence shows that hallucinated package names are predictable and recur across models and prompts, enabling systematic squatting on high-probability ``ghost'' identifiers~\cite{spracklen2025package}.

    \item \textbf{Semantic Masquerading:}
    Discovery mechanisms commonly rely on semantic similarity between user intent and tool descriptions. By adversarially crafting metadata to maximize textual overlap with common queries, attackers can manipulate retrieval rankings and displace legitimate tools. This breaks semantic binding by associating a correct intent with an incorrect capability. Benchmarks show that even minor perturbations to tool descriptions can significantly degrade selection accuracy for strong models~\cite{ye2024rotbench}, and tool-selection studies explicitly identify description noise as a practical misdirection channel~\cite{ye2402toolsword}.
\end{enumerate}

\subsection{Phase II: Attacks on Implementation (Load \texttt{Tool\_ID} $\rightarrow$ Runtime Code)}

After a tool identifier is resolved, the Implementation phase fetches and instantiates executable logic:
\begin{equation}
\text{Load}(\texttt{Tool\_ID}) \rightarrow \text{Runtime Code}.
\end{equation}
Unlike Discovery attacks, which manipulate selection, Implementation attacks preserve correct tool choice while corrupting the code that executes.
\begin{enumerate}
    \item \textbf{Hidden Backdoors and Malicious Extensions:}
A tool may implement its advertised functionality under benign conditions while embedding trigger-based logic that activates malicious behavior in specific contexts. Such backdoors allow agents to maintain high apparent utility while silently executing unintended actions, compromising implementation integrity without visible deviation in reasoning traces. This failure mode has been demonstrated in backdoored agent backends that achieve high attack success rates while preserving normal task performance~\cite{wang2024badagent}.

    \item \textbf{Transitive Dependency Exploitation:}
Implementation integrity may also be compromised indirectly through dependency resolution. Tool setup often triggers installation of auxiliary packages, and common package managers permit arbitrary code execution during installation. In autonomous workflows, agents routinely modify their environments to complete tasks, treating dependency installation as a functional step rather than a security-sensitive operation~\cite{fang2024llm}. As a result, selecting a benign tool may implicitly execute malicious transitive dependencies before the intended tool is ever invoked.

\end{enumerate}


\subsection{Phase III: Attacks on the Invocation Boundary}

The Invocation phase governs how a selected tool is executed at runtime. Unlike earlier phases, these attacks arise from failures in constraining \emph{how} a tool is used rather than which tool is selected or what code is loaded.

We model invocation as:
\begin{equation}
\text{Invoke}(T, \theta, \mathcal{C}) \rightarrow (o, \Delta),
\end{equation}
where $\theta$ denotes invocation arguments and $\mathcal{C}$ the execution context (e.g., credentials and accessible resources). Phase~III attacks aim to induce unauthorized or unintended side effects $\Delta$, or to expose sensitive elements of $\mathcal{C}$, even when $T$ is legitimate.

\begin{enumerate}
   \item \textbf{Over-Privileged Invocation:}

Agents are frequently provisioned with broad permissions and long-lived credentials to ensure task completion. When invocation lacks capability-based constraints, adversaries can induce a confused-deputy–like failure in which the agent applies its authority to actions not justified by the originating intent.

This phenomenon has been formalized in recent work on privilege escalation in LLM-based multi-agent systems, where natural-language communication channels allow low-privilege components to trigger high-privilege tool execution~\cite{ji2026taming}. Because authority usage is driven by the agent’s internal reasoning rather than explicit human approval, such failures are non-interactive and may be difficult to detect or reverse once committed.

Empirical studies further show that indirect prompt injection can blur the boundary between data and instructions, enabling authority hijacking in tool-integrated agents~\cite{abdelnabi2023indirect}. This threat is operationalized in \textit{InjecAgent}, where agents can be induced to perform unauthorized transfers or data exfiltration when invocation is not mediated by explicit permission checks~\cite{zhan2403injecagent}. Additional evaluations demonstrate that agents frequently misjudge side-effect boundaries even for syntactically valid tool calls, resulting in severe consequences such as irreversible data deletion~\cite{ruan2024toolemu}.

   \item \textbf{Argument Injection and Logical Boundary Breaches.}

Even when a tool $T$ is correctly selected and invoked under valid credentials, adversarial manipulation of invocation parameters $\theta$ can induce side effects that exceed intended task semantics. In this setting, the capability itself is legitimate, but the mapping between arguments and permissible state transitions is insufficiently constrained.

This failure mode is analogous to application-layer exploits such as Server-Side Request Forgery (SSRF), where a trusted intermediary relays attacker-controlled inputs into a privileged execution context. Empirical evidence shows that indirect prompt injection can reliably steer downstream API arguments without altering tool identity~\cite{kong2024injectbench}, and recent work demonstrates how prompt injection increasingly merges with classical web vulnerabilities to exploit application-layer weaknesses~\cite{mchugh2025prompt}. Industry analyses further identify Cross-Plugin Request Forgery (CPRF), where injected content causes agents to silently generate malicious parameters for connected tools.

Unlike over-privileged invocation, which violates \emph{authority bounding}, argument injection violates \emph{semantic constraint binding} at the invocation boundary. From a supply-chain perspective, the tool and authority remain authentic, yet logical boundary enforcement between intent, argument space, and external side effects fails.
\end{enumerate}

\section{The Viral Agent: Closing the Loop}
\label{sec:viral_agent}

Prior sections treat the Data and Tool supply chains separately. A defining risk of agentic systems is \emph{autonomous replication}: when data ingestion (Discovery) is coupled with tool execution (Invocation), agent outputs can re-enter the environment as future inputs, closing a loop between consumption and contamination.

\subsection{The Self-Replicating Cycle}

We define a \textbf{Viral Agent Loop} as the recursive process where agent $\text{Ag}_A$ consumes a payload $P$, triggers a side effect $\Delta$, and $\Delta$ is later retrieved by another agent $\text{Ag}_B$ as input:
\begin{equation}
\text{Invoke}(\text{Ag}_A, P) \xrightarrow{\;\text{writes}\;} \Delta_{\text{env}}
\;\;\wedge\;\;
\text{Retrieve}(\text{Ag}_B) \ni \Delta_{\text{env}} \supseteq P.
\end{equation}
Thus, $\Delta_{\text{env}}$ becomes part of $\text{Ag}_B$'s effective context, turning agents into carriers. Unlike traditional malware, propagation can occur without code vulnerabilities, exploiting instruction-following under legitimate tool permissions.

\subsection{Propagation Vectors and Evidence}
\begin{enumerate}
    \item \textbf{Generative Worms.}
    \textit{Morris II}~\cite{cohen2024morris} demonstrates prompt-based self-replication across GenAI agents: an email agent can be induced to exfiltrate context and forward the payload via authorized communication tools. This shifts propagation from \emph{syntactic} exploits to \emph{semantic} compliance, making patching insufficient.

    \item \textbf{Persistent Ecosystem Contamination.}
    If agents can write to shared repositories (e.g., wikis, version control), poisoned outputs may be uploaded and later re-ingested through retrieval, enabling persistence across sessions and agents and degrading shared knowledge integrity.

    \item \textbf{Topological Shift: Pipeline $\rightarrow$ Cycle.}
    Traditional supply chains resemble DAGs; agentic systems form \emph{cycles} because side effects $\Delta$ can become future inputs. This collapses the supplier/consumer boundary and invalidates defenses that assume acyclic dependency resolution.
\end{enumerate}

\section{Systematization of Existing Defenses}
\label{sec:defenses}
Prior work emphasizes static LLM alignment \cite{wang2024supplychain}. However, dynamic Agentic Supply Chains require a \textit{Zero Trust Runtime} architecture. We systematize existing defenses across five runtime layers (Perception, Memory, Resolution, Implementation, Invocation), mapped to our threat model in Table~\ref{tab:defense_matrix}.

\begin{table}[htbp]
\caption{Systematization of Existing Agentic Supply Chain Defenses}
\vspace{-2mm}
\begin{center}
\begin{tabular}{|p{0.22\linewidth}|p{0.28\linewidth}|p{0.38\linewidth}|}
\hline
\textbf{Supply Chain Component} & \textbf{Targeted Attack Vector} & \textbf{Existing Defensive Paradigms} \\
\hline
\textbf{Data (Perception)} & Direct \& Indirect Prompt Injection & Instruction Hierarchy \cite{wallace2024instructionhierarchytrainingllms}, Intent Verification \cite{jia2024taskshieldenforcingtask, kang2025mitigatingindirectpromptinjection} \\
\hline
\textbf{Data (Memory)} & RAG \& Agent-State Poisoning & Statistical Filtering \cite{ragdefender2025rescuing}, Audited Memory Writes \cite{amemguard2025proactive} \\
\hline
\textbf{Tool (Phase I)} & Hallucination Squatting & Static Registry Allowlists \\
\hline
\textbf{Tool (Phase II)} & Dependency Pollution & Signed SBOMs, SLSA Frameworks \cite{slsa2024framework} \\
\hline
\textbf{Tool (Phase III)} & Privilege Escalation & Scoped Permissions (e.g., MCP \cite{anthropic2024mcp}), Multi-stage Arbiters \cite{xing2026mcpguard} \\
\hline
\end{tabular}
\label{tab:defense_matrix}
\end{center}
\vspace{-4mm}
\end{table}

While the paradigms in Table~\ref{tab:defense_matrix} offer valuable localized mitigations, they fundamentally inherit assumptions from the \textit{Static Model Supply Chain}. By evaluating these defenses through their respective supply chain components, we expose why they remain structurally insufficient against the MitE adversary.

\subsection{Securing the Data Supply Chain}
Defenses at this layer must (i) structurally separate operator instructions from passive data, and (ii) police memory integrity.

\paragraph{Perception Layer (Instruction \& Intent)}
Frameworks such as the \textbf{Instruction Hierarchy}~\cite{wallace2024instructionhierarchytrainingllms} and \textbf{Intent Verification}~\cite{jia2024taskshieldenforcingtask, kang2025mitigatingindirectpromptinjection} encapsulate untrusted external data within rigid structural delimiters. However, this approach underestimates the \textit{epistemic uncertainty} inherent in In-Context Learning. Because agents continuously upcast passive data into active context, adversaries can exploit the Prompt-Data Isomorphism to bypass syntactic delimiters using purely semantic payloads. Consequently, generative worms~\cite{cohen2024morris} can still induce malicious behavior without violating the prescribed structural hierarchy.

\paragraph{Memory Layer (Persistence)}
To prevent persistent knowledge corruption, mechanisms rely on \textbf{Statistical Filtering}~\cite{ragdefender2025rescuing} or \textbf{Audited Memory Writes}~\cite{amemguard2025proactive}. Unfortunately, empirical evidence demonstrates that statistical anomaly detection falls short against structural graph poisoning~\cite{liang2025graphrag}. When an adversary subtly manipulates relational edges in an agent's long-term memory or external RAG corpus, the retrieved text appears statistically benign but logically steers the agent into a compromised operational state. 

\subsection{Securing the Tool Supply Chain}
Tool defenses operate sequentially across the execution pipeline to restore identity, integrity, and authority.

\paragraph{Resolution and Implementation (Phases I \& II)}
To counter hallucination squatting and dependency pollution, current proposals advocate for \textbf{Static Registry Allowlists}~\cite{spracklen2025package} and verifiable provenance via \textbf{SBOMs}~\cite{slsa2024framework}. While these measures guarantee identity and implementation integrity, they are blind to the agent's semantic intent. An agent manipulated by MitE will simply utilize a cryptographically verified, perfectly legitimate tool to execute an unauthorized, malicious action.

\paragraph{Phase III: Invocation Integrity \& Semantic Firewalls}
This phase represents the final line of defense before side effects occur. Recognizing that standard OS-level permissions cannot detect semantic misuse (e.g., a legally authorized but contextually malicious file deletion), recent architectures introduce \textbf{Semantic Firewalls} (e.g., Vigil~\cite{lin2026vigil}, MCP-Guard~\cite{xing2026mcpguard}) and ephemeral capabilities (e.g., MCP~\cite{anthropic2024mcp}). Yet, this ``AI-guarding-AI'' paradigm suffers from the exact same fundamental vulnerability: the arbiter model must process the identical tainted context, thus inheriting the vulnerability to prompt injection. To bridge the translation gap between benign user intent and hazardous execution, we argue for a \textbf{Defense-in-Depth} approach combining AI oversight with structural bounding:

\section{The Zero-Trust Agentic Runtime Architecture}
\label{sec:zero_trust}



Collectively, the localized defenses in Section 6 treat execution as an acyclic pipeline, applying static patches that fail to neutralize the \textit{Viral Agent Loop}. They treat epistemic uncertainty and stochastic capabilities as isolated ML safety issues rather than interconnected supply chain vulnerabilities. To effectively counter the MitE adversary, the security paradigm must shift from the Static Model Supply Chain to a \textbf{Zero-Trust Dynamic Agentic Runtime Architecture}. Securing real-time data consumption, tool execution, and cyclic propagation is predicated on three core imperatives:

\subsection{Deterministic Capability Binding}
The reliance on LLMs to probabilistically "guess" tool identifiers via semantic similarity is structurally insecure and constitutes the root cause of Phase I and Phase II tool supply chain attacks. A Zero-Trust Runtime must enforce \textbf{Deterministic Capability Binding}. We advocate for the deployment of \textbf{Cryptographically Bound Registries} that verify the semantic integrity of tool providers. Under this architecture, tool execution is bounded strictly by cryptographic provenance rather than semantic likelihood. The translation from natural-language intent to executable capability must be mediated by a verified resolution engine that mathematically guarantees identity integrity, eliminating the "Hallucination Gap" entirely.

\subsection{Neuro-Symbolic Information Flow Control}
Current static analysis techniques are insufficient for dynamic agents because they fail to account for the \textit{Viral Agent Loop}, where outputs re-enter the environment as tainted context. To prevent self-propagating agentic worms, we propose \textbf{Runtime Taint Analysis} engineered for neural systems. 

In this framework, untrusted external inputs (e.g., retrieved web text) are strictly tagged as \texttt{TAINTED}. The runtime must track this taint through the non-deterministic transformations of the LLM's reasoning chain. If a tainted reasoning trace attempts to invoke a write-privileged sink (e.g., \texttt{send\_email} or \texttt{git\_push}), the execution is blocked pending explicit sanitization or human-in-the-loop approval. Because neural networks can wash standard metadata tags through paraphrasing, this flow control must be underpinned by \textbf{Cryptographic Provenance Ledgers}~\cite{li2025towardtrustworthyagentic}, ensuring an immutable record of data lineage from perception to action.

\subsection{The Auditor-Worker Architecture (Semantic Firewalls)}
Standard OS-level sandboxing cannot prevent logical boundary breaches (Phase III attacks) because the runtime cannot distinguish between a legally authorized but contextually malicious action (e.g., deleting a critical database file at the behest of an injected prompt) and a benign task. A single model cannot simultaneously optimize for both stochastic helpfulness and deterministic safety.

We propose the \textbf{Auditor-Worker Architecture}. This paradigm structurally decouples execution from oversight by placing an isolated \textbf{Supervisor Model} as an inline \textbf{Semantic Firewall}. Operating via speculative execution (extending concepts from preliminary frameworks like \textit{Vigil}~\cite{lin2026vigil}, \textit{AgentGuard}~\cite{chen2025agentguard}, and multi-stage cognitive arbiters like \textit{MCP-Guard}~\cite{xing2026mcpguard}), the Supervisor analyzes the proposed tool call, its parameters, and the lineage of the prompt \textit{before} the action is committed to the environment. This hard structural split bridges the translation gap between benign user intent and hazardous execution, enforcing least-privilege constraints at the semantic layer.


\section{Conclusion and Future Directions}
\label{sec:conclusion}

This Systematization of Knowledge formalizes the topological shift from static software security to the \textit{Dynamic Agentic Runtime Supply Chain (DRSC)}. We defined the \textit{Man-in-the-Environment (MitE)} adversary, who exploits \textit{Prompt-Data Isomorphism} to weaponize an agent's perception and tool usage. Crucially, we identified the \textit{Viral Agent Loop}, where agents become vectors for self-propagating generative worms without underlying code vulnerabilities. 




To secure the dynamic agentic runtime against the Man-in-the-Environment (MitE) adversary, future research must address three interconnected imperatives. First, we require \textbf{Realistic Evaluation Benchmarks} that move beyond static simulations with closed-world assumptions and episodic amnesia~\cite{zhan2403injecagent,debenedetti2024agentdojodynamicenvironmentevaluate}. Future frameworks must evaluate how agents dynamically locate and trust tools in open-world, hostile environments to capture persistent memory poisoning and cyclic viral propagation. Second, the development of \textbf{Robust Capability Protocols} is essential; while standardization efforts like MCP~\cite{anthropic2024mcp} are promising, their reliance on semantic resolution leaves them vulnerable to description injection~\cite{wang2025mpmapreferencemanipulationattack}, necessitating mathematically verifiable, Name-Squatting Resilient Registries. Finally, mitigating epistemic uncertainty requires \textbf{Neuro-Symbolic Taint Tracking}---developing reliable mechanisms to maintain cryptographic provenance tags across the non-deterministic, stochastic abstractions of transformer layers, even when syntactic representations change.

Ultimately, in the Agentic Supply Chain, context \textit{is} code. Until security architectures fully embrace this prompt-data isomorphism, autonomous agents will remain fundamentally vulnerable operators. 



\raggedbottom

\subsubsection*{Reproducibility Statement}
We confirm that the code and datasets needed to reproduce the experiments are available in the supplementary material. Detailed proofs for the theoretical claims are provided in Appendix A. The taxonomy presented relies on publicly available attack demonstrations cited throughout the text.

\subsubsection*{Ethics Statement}
This work focuses on the security vulnerabilities of autonomous agents. While the described attack vectors could potentially be misused, our goal is to systemize knowledge to facilitate better defense strategies. We have adhered to standard responsible disclosure practices where applicable.

\bibliographystyle{iclr2026_conference}
\bibliography{iclr2026_conference}

@phdthesis{kong2024injectbench,
  title={InjectBench: An Indirect Prompt Injection Benchmarking Framework},
  author={Kong, Nicholas Ka-Shing},
  year={2024},
  school={Virginia Tech}
}

@article{mchugh2025prompt,
  title={Prompt injection 2.0: hybrid AI threats},
  author={McHugh, Jeremy and {\v{S}}ekrst, Kristina and Cefalu, Jon},
  journal={arXiv preprint arXiv:2507.13169},
  year={2025}
}

@article{ji2026taming,
  title={Taming Various Privilege Escalation in LLM-Based Agent Systems: A Mandatory Access Control Framework},
  author={Ji, Zimo and Wu, Daoyuan and Jiang, Wenyuan and Ma, Pingchuan and Li, Zongjie and Gao, Yudong and Wang, Shuai and Li, Yingjiu},
  journal={arXiv preprint arXiv:2601.11893},
  year={2026}
}

@misc{wang2023voyager,
      title={Voyager: An Open-Ended Embodied Agent with Large Language Models}, 
      author={Guanzhi Wang and Yuqi Xie and Yunfan Jiang and Ajay Mandlekar and Chaowei Xiao and Yuke Zhu and Linxi Fan and Anima Anandkumar},
      year={2023},
      eprint={2305.16291},
      archivePrefix={arXiv},
      primaryClass={cs.AI},
}

@misc{wang2024supplychain,
      title={Large Language Model Supply Chain: A Research Agenda}, 
      author={Shenao Wang and Yanjie Zhao and Xinyi Hou and Haoyu Wang},
      year={2024},
      eprint={2404.12736},
      archivePrefix={arXiv},
      primaryClass={cs.SE},
}

@misc{xi2025rise,
      title={The Rise and Potential of Large Language Model Based Agents: A Survey}, 
      author={Zhiheng Xi and others},
      year={2023},
      eprint={2309.07864},
      archivePrefix={arXiv},
      primaryClass={cs.AI},
}

@misc{abdelnabi2023indirect,
      title={Not what you've signed up for: Compromising Real-World LLM-Integrated Applications with Indirect Prompt Injection}, 
      author={Kai Greshake and Sahar Abdelnabi and Shailesh Mishra and Christoph Endres and Thorsten Holz and Mario Fritz},
      year={2023},
      eprint={2302.12173},
      archivePrefix={arXiv},
      primaryClass={cs.CR},
}

@misc{wang2024badagent,
      title={BadAgent: Inserting and Activating Backdoor Attacks in LLM Agents}, 
      author={Yifei Wang and Dizhan Xue and Shengjie Zhang and Shengsheng Qian},
      year={2024},
      eprint={2406.03007},
      archivePrefix={arXiv},
      primaryClass={cs.CL},
}

@misc{fang2024llm,
      title={LLM Agents can Autonomously Hack Websites}, 
      author={Richard Fang and Rohan Bindu and Akul Gupta and Qiusi Zhan and Daniel Kang},
      year={2024},
      eprint={2402.06664},
      archivePrefix={arXiv},
      primaryClass={cs.CR},
       
}

@misc{cohen2024morris,
      title={Here Comes The AI Worm: Unleashing Zero-click Worms that Target GenAI-Powered Applications}, 
      author={Stav Cohen and Ron Bitton and Ben Nassi},
      year={2025},
      eprint={2403.02817},
      archivePrefix={arXiv},
      primaryClass={cs.CR},
       
}

@misc{zou2024poisonedrag,
      title={PoisonedRAG: Knowledge Corruption Attacks to Retrieval-Augmented Generation of Large Language Models}, 
      author={Wei Zou and Runpeng Geng and Binghui Wang and Jinyuan Jia},
      year={2024},
      eprint={2402.07867},
      archivePrefix={arXiv},
      primaryClass={cs.CR},
       
}

@misc{yao2025minja,
      title={Memory Injection Attacks on LLM Agents via Query-Only Interaction}, 
      author={Shen Dong and Shaochen Xu and Pengfei He and Yige Li and Jiliang Tang and Tianming Liu and Hui Liu and Zhen Xiang},
      year={2026},
      eprint={2503.03704},
      archivePrefix={arXiv},
      primaryClass={cs.LG},
       
}

@misc{chen2024agentpoison,
      title={AgentPoison: Red-teaming LLM Agents via Poisoning Memory or Knowledge Bases}, 
      author={Zhaorun Chen and Zhen Xiang and Chaowei Xiao and Dawn Song and Bo Li},
      year={2024},
      eprint={2407.12784},
      archivePrefix={arXiv},
      primaryClass={cs.LG},
       
}

@misc{bagdasaryan2023abusingimagessoundsindirect,
      title={Abusing Images and Sounds for Indirect Instruction Injection in Multi-Modal {LLMs}}, 
      author={Eugene Bagdasaryan and Tsung-Yin Hsieh and Ben Nassi and Vitaly Shmatikov},
      year={2023},
      eprint={2307.10490},
      archivePrefix={arXiv},
      primaryClass={cs.CR},
       
}

@inproceedings{anil2024manyshot,
  title={Many-shot Jailbreaking},
  author={Anil and others},
  booktitle={Proc. of NeurIPS},
  year={2024},
  url={https://www.anthropic.com/research/many-shot-jailbreaking}
}

@misc{zhong2023poisoning,
      title={Poisoning Retrieval Corpora by Injecting Adversarial Passages}, 
      author={Zexuan Zhong and Ziqing Huang and Alexander Wettig and Danqi Chen},
      year={2023},
      eprint={2310.19156},
      archivePrefix={arXiv},
      primaryClass={cs.CL},
       
}

@misc{an2025ipiguard,
      title={IPIGuard: A Novel Tool Dependency Graph-Based Defense Against Indirect Prompt Injection in LLM Agents}, 
      author={Hengyu An and Jinghuai Zhang and Tianyu Du and Chunyi Zhou and Qingming Li and Tao Lin and Shouling Ji},
      year={2025},
      eprint={2508.15310},
      archivePrefix={arXiv},
      primaryClass={cs.CR},
       
}

@inproceedings{ruan2024toolemu,
  title={ToolEmu: Identifying the Risks of {LM} Agents with an {LM}-Emulated Sandbox},
  author={Ruan, Yangjun and others},
  booktitle={Proc. of ICLR},
  year={2024}
}

@inproceedings{ye2024rotbench,
  title={Rotbench: A multi-level benchmark for evaluating the robustness of large language models in tool learning},
  author={Ye, Junjie and others},
  booktitle={Proc. of EMNLP},
  pages={313--333},
  year={2024}
}

@misc{ye2402toolsword,
      title={ToolSword: Unveiling Safety Issues of Large Language Models in Tool Learning Across Three Stages}, 
      author={Junjie Ye and others},
      year={2024},
      eprint={2402.10753},
      archivePrefix={arXiv},
      primaryClass={cs.CL},
       
}

@article{lee2024prompt,
  title={Prompt Infection: LLM-to-LLM Prompt Injection within Multi-Agent Systems},
  author={Lee, Donghyun and Tiwari, Mo},
  journal={arXiv preprint arXiv:2410.07283},
  year={2024}
}

@misc{wallace2024instructionhierarchytrainingllms,
      title={The Instruction Hierarchy: Training LLMs to Prioritize Privileged Instructions}, 
      author={Eric Wallace and Kai Xiao and Reimar Leike and Lilian Weng and Johannes Heidecke and Alex Beutel},
      year={2024},
      eprint={2404.13208},
      archivePrefix={arXiv},
      primaryClass={cs.CR},
       
}

@misc{kang2025mitigatingindirectpromptinjection,
      title={Mitigating Indirect Prompt Injection via Instruction-Following Intent Analysis}, 
      author={Mintong Kang and Chong Xiang and Sanjay Kariyappa and Chaowei Xiao and Bo Li and Edward Suh},
      year={2025},
      eprint={2512.00966},
      archivePrefix={arXiv},
      primaryClass={cs.CR},
       
}

@misc{lin2026vigil,
      title={VIGIL: Defending LLM Agents Against Tool Stream Injection via Verify-Before-Commit}, 
      author={Junda Lin and Zhaomeng Zhou and Zhi Zheng and Shuochen Liu and Tong Xu and Yong Chen and Enhong Chen},
      year={2026},
      eprint={2601.05755},
      archivePrefix={arXiv},
      primaryClass={cs.CR},
       
}

@misc{jia2024taskshieldenforcingtask,
      title={The Task Shield: Enforcing Task Alignment to Defend Against Indirect Prompt Injection in LLM Agents}, 
      author={Feiran Jia and Tong Wu and Xin Qin and Anna Squicciarini},
      year={2024},
      eprint={2412.16682},
      archivePrefix={arXiv},
      primaryClass={cs.CR},
       
}

@misc{li2025towardtrustworthyagentic,
      title={Toward Trustworthy Agentic AI: A Multimodal Framework for Preventing Prompt Injection Attacks}, 
      author={Toqeer Ali Syed and Mishal Ateeq Almutairi and Mahmoud Abdel Moaty},
      year={2025},
      eprint={2512.23557},
      archivePrefix={arXiv},
      primaryClass={cs.CR},
       
}

@misc{ragdefender2025rescuing,
      title={Rescuing the Unpoisoned: Efficient Defense against Knowledge Corruption Attacks on RAG Systems}, 
      author={Minseok Kim and Hankook Lee and Hyungjoon Koo},
      year={2025},
      eprint={2511.01268},
      archivePrefix={arXiv},
      primaryClass={cs.CR},
       
}

@misc{amemguard2025proactive,
      title={A-MemGuard: A Proactive Defense Framework for LLM-Based Agent Memory}, 
      author={Qianshan Wei and Tengchao Yang and Yaochen Wang and Xinfeng Li and Lijun Li and Zhenfei Yin and Yi Zhan and Thorsten Holz and Zhiqiang Lin and XiaoFeng Wang},
      year={2025},
      eprint={2510.02373},
      archivePrefix={arXiv},
      primaryClass={cs.CR},
       
}

@misc{slsa2024framework,
  title={Supply-chain Levels for Software Artifacts (SLSA) v1.0},
  author={OpenSSF},
  year={2024},
  note={\url{https://slsa.dev/}}
}

@misc{chen2025agentguard,
      title={AgentGuard: Repurposing Agentic Orchestrator for Safety Evaluation of Tool Orchestration}, 
      author={Jizhou Chen and Samuel Lee Cong},
      year={2025},
      eprint={2502.09809},
      archivePrefix={arXiv},
      primaryClass={cs.CR},
       
}

@misc{anthropic2024mcp,
  title={The Model Context Protocol (MCP): Standardizing AI Context},
  author={Anthropic},
  year={2024},
  note={\url{https://www.anthropic.com/news/model-context-protocol}}
}

@inproceedings{spracklen2025package,
  title={We Have a Package for You! A Comprehensive Analysis of Package Hallucinations by Code Generating LLMs},
  author={Spracklen, Joseph and others},
  booktitle={34th USENIX Security Symposium (USENIX Security 25)},
  year={2025},
  note={arXiv:2406.10279}
}

@misc{dong2023survey,
      title={A Survey on In-context Learning}, 
      author={Qingxiu Dong and others},
      year={2024},
      eprint={2301.00234},
      archivePrefix={arXiv},
      primaryClass={cs.CL},
       
}

@misc{weng2023memory,
      title={A Survey on the Memory Mechanism of Large Language Model based Agents}, 
      author={Zeyu Zhang and Xiaohe Bo and Chen Ma and Rui Li and Xu Chen and Quanyu Dai and Jieming Zhu and Zhenhua Dong and Ji-Rong Wen},
      year={2024},
      eprint={2404.13501},
      archivePrefix={arXiv},
      primaryClass={cs.AI},
       
}

@misc{wang2025mpmapreferencemanipulationattack,
      title={MPMA: Preference Manipulation Attack Against Model Context Protocol}, 
      author={Zihan Wang and Rui Zhang and Yu Liu and Wenshu Fan and Wenbo Jiang and Qingchuan Zhao and Hongwei Li and Guowen Xu},
      year={2025},
      eprint={2505.11154},
      archivePrefix={arXiv},
      primaryClass={cs.CR},
       
}

@misc{zhan2403injecagent,
      title={InjecAgent: Benchmarking Indirect Prompt Injections in Tool-Integrated Large Language Model Agents}, 
      author={Qiusi Zhan and Zhixiang Liang and Zifan Ying and Daniel Kang},
      year={2024},
      eprint={2403.02691},
      archivePrefix={arXiv},
      primaryClass={cs.CL},
       
}

@misc{debenedetti2024agentdojodynamicenvironmentevaluate,
      title={AgentDojo: A Dynamic Environment to Evaluate Prompt Injection Attacks and Defenses for LLM Agents}, 
      author={Edoardo Debenedetti and Jie Zhang and Mislav Balunović and Luca Beurer-Kellner and Marc Fischer and Florian Tramèr},
      year={2024},
      eprint={2406.13352},
      archivePrefix={arXiv},
      primaryClass={cs.CR},
       
}

@misc{liang2025graphrag,
      title={GraphRAG under Fire}, 
      author={Jiacheng Liang and Yuhui Wang and Changjiang Li and Rongyi Zhu and Tanqiu Jiang and Neil Gong and Ting Wang},
      year={2025},
      eprint={2501.14050},
      archivePrefix={arXiv},
      primaryClass={cs.LG},
       
}

@misc{xing2026mcpguard,
      title={MCP-Guard: A Multi-Stage Defense-in-Depth Framework for Securing Model Context Protocol in Agentic AI}, 
      author={Wenpeng Xing and others},
      year={2026},
      eprint={2508.10991},
      archivePrefix={arXiv},
      primaryClass={cs.CR},
       
}

@misc{zhao2025patronusidentifyingmitigatingtransferable,
      title={Patronus: Identifying and Mitigating Transferable Backdoors in Pre-trained Language Models}, 
      author={Tianhang Zhao and Wei Du and Haodong Zhao and Sufeng Duan and Gongshen Liu},
      year={2025},
      eprint={2512.06899},
      archivePrefix={arXiv},
      primaryClass={cs.CR},
      url={https://arxiv.org/abs/2512.06899}, 
}

@article{Liu_2026,
title={Security of Large Model-based Agents: A Survey on Adversarial, Poisoning, and Backdoor Attacks},
url={http://dx.doi.org/10.36227/techrxiv.177006506.61959855/v1},
DOI={10.36227/techrxiv.177006506.61959855/v1},
publisher={Institute of Electrical and Electronics Engineers (IEEE)},
author={Liu, Xinyun and Cheng, Zelei and Zhao, Haodong and Xu, Ronghua},
year={2026},
month=feb }

\newpage
\appendix

\section{LLM Usage Statement}
The authors acknowledge the use of Gemini to refine the
clarity and grammar of the text. The authors reviewed and
revised the output and take full responsibility for the content of this article.
\end{document}